\documentclass[12pt]{iopart}

\begin{document}

\title[ ]{Primary photons, boson star and phase transition}

\author{G.A. Kozlov}

\address{JINR, BLTP, Joliot-Curie st. 6, Dubna, 141980, Russia}
\ead{kozlov@jinr.ru}
\vspace{10pt}
\begin{indented}
\item[]
\end{indented}

\begin{abstract}

We analyse the possibility that the dark matter candidate is from the approximate scale symmetry theory of the hidden scalar sector. The study includes the warm dark matter scenario and the Bose-Einstein condensation which may lead to the scalar  boson stars (BS) giving rise to direct detection through the observation of the primary (direct) photons.
The dynamical system of the scalar particles, the dilatons, at finite temperature and chemical potential is considered.
The fluctuation of the particle density increases sharply within the increasing of the temperature. When the phase transition approaches,  the fluctuation of the particle density has the non-monotonous rising when the ground state of the relative chemical potential tends to the critical value equal to one accompanying by the infinite number of particles. 
Our results suggest that the phase transition in the BS may be identified through the fluctuation  in yield of primary photons induced directly by the conformal anomaly. The fluctuation rate of the photons grows up intensively in the infra-red to become very large at the phase transition.    
%\end{abstract}

%\keywords{Boson star; phase transition; conformal anomaly.}
\end{abstract}

%\ccode{PACS Nos.: 12.60.Fr, 11.10.Ef, 11.15.Ex.}

\section{Introduction}	

The spontaneous breaking symmetry has a major role in particle physics and cosmology where the phase transitions (PT) can occur at some extreme conditions (e.g., high enough temperature, the high value of the chemical potential, etc.). The thermal corrections to the Higgs potential may restore the electroweak (EW) gauge symmetry in the early Universe. 
%For the mass of the Higgs boson the PT is a smooth crossover rather than the true PT [1-3].
In the dark (hidden) sector, the PT corresponds to an initial thermal state which is invariant under the conformal group $G$. This state has a nontrivial couplings to the other states being the physical matter fields. These fields do not possess the invariance under $G$, but they are invariant under the chiral symmetry group triggered by $G$. For the physical mass of the Higgs boson, the standard model (SM) symmetry restoration with the PT is a cross-over rather than the true PT [1]. A cross-over means that the early Universe evolved smoothly from an unbroken symmetric to broken phase where no significant deviations from the thermal equilibrium are emerged. However, the cross-over can influence the other processes, e.g., the formation of the gravitationally approximate stable bound states composed of the dark matter (DM) fields under the repulsive forces between them.
%If the weak and the EW scales are embedded in the SM, one can apply the complete formalism of the high-temperature weak and EW transition between the low-energy broken phase where the scalar field has has a non-vanishing %expectation value, and a high-temperature symmetric phase. 
In the early Universe at high temperatures, the particles exhibit an asymptotic freedom that means there is the gas of almost free light massive and/or massless particles.  All the scales in particle physics and cosmology  are the subjects of light scalar fields which are the DM candidates. 
%If the weak and the EW scales are embedded in the SM, one can apply the complete formalism of the high-temperature weak and EW transition between the low-energy broken phase where the scalar field has a non-vanishing expectation value, and a high-temperature symmetric phase. 
If the self-interactions in scalar hidden sector are repulsive the (very) massive boson stars (BS) as the bound states of the scalar fields can be emerged inward gravitational force and is balanced by the repulsive self-interaction [2].
In theories beyond the SM, the introducing the additional scalar fields can result in the potential having the different temperature-dependent behaviour from that of the SM. 
%The quantum effects, e.g., the fluctuations of gluons, break the scale invariance. 
As a result, 
when the temperature of the hidden scalar sector is on order of the strong coupling scale $\sim \Lambda$, 
%the light particles with spin 0  are expected to be emerged. In particular, 
the scalar particles can lock up into colour spin 0 neutral states, the "glueballs", whose mass is of the order of $\Lambda$ [3]. 
 The scalar dilaton in the form of the glueball is a well-motivated example coming from a pure Yang-Mills hidden sector, which locks up into the bound state of two gluons in the early Universe. 
 %The glueball DM may condense into the boson stars and be observed by gravitational  lensing effect.  
There are expectations in the glueballs existence as hadronic (colourless) states in their own right as simpler structures than the more conventional hadronic states involving quarks [4-6].  
The glueball DM may condense into the boson stars and be observed by gravitational  lensing effect.  
%The lightest hidden scalar fields with the mass $ < 10^{4}$ TeV could be the candidate for DM. 
These lightest hidden scalars could be cosmologically long lived with the lifetime  $\sim 10^{17}$ sec (approximately the age of the Universe) giving by the decay width of the scalar into two gravitons if the mass of the scalar $ < 10^{4}$ TeV [7].

%All scales in particle physics and cosmology  are the subjects of the light scalar fields.
The critical phenomena if occurred may be considered through the quantum PT with  the Bose-Einstein condensation (BEC) of the scalar  field. In this case, the condensation takes place in a single zero mode that suggests the breaking of conformal symmetry. 
The $SU(3)$ gauge theory studied on the lattice with $N_{f} = 8$ Dirac fermions in fundamental representation show an evidence  of a light scalar singlet [8]. The latter might be a dilaton in the effective field theory [9-11] well before the recent activities in which the dilaton, a pseudo-Goldstone boson of the spontaneous breaking of conformal symmetry, provide, in particular, a portal between the DM and the visible sector of matter [12-14].
%The properties of quarks and gluons accompanying by the scalar condensate at high both temperature and baryon chemical potential are the keys for understanding the evolution of the Universe.
The dynamics and the properties of the dilatons (and their decays) at the temperature and the chemical potential around their critical values may be the keys to understanding the evolution of the Universe.
%There are  also more or less the same pattern for experiments with heavy ion collisions at early times. 
%One can imply the existence of the Goldstone-like modes: the scale symmetry is broken explicitly resulting with an appearance of the dilaton in the spectrum, accompanying by $\pi$-mesons as a consequence of the chiral symmetry %breaking. 
%In physics aiming the search for CP and PT at high temperatures in strongly dense (nuclear) matter %the effects of explicit breaking of scale symmetry may have the dominant feature. The latter seems to %trigger the confinement process in strong interacting matter through the spontaneous breaking of %chiral symmetry.
The existence of the strong interactions at finite temperature is a very good example for the theory that possesses the conformal invariance at the classical level because there are no dimensional parameters associated with the classical formulation of the theory.
The quantum effects, e.g. the gluon fluctuations, break the conformal (scale) invariance. 
This assumes that all the processes are governed by the conformal anomaly (CA) resulting from the running coupling constant $\alpha_{s}$. 
%From this $S_{\mu}$ is already non-conserved in the theory containing, e.g., the gluon and quark %degrees of freedom (d.o.f.): 
The theory becomes conformal in the infra-red (IR)  with the non-trivial solution for $\alpha_{s}$ as the ratio between two first coefficients of the $\beta$-function, 
$\alpha_{s}^{\star} \sim - b_{0}/b_{1}$ (the IR fixed point (IRFP) or the Banks-Zaks [15] conformal point), in the perturbative domain if $b_{0} = (11 N_{c}- 2 N_{f})/3$ is small ($N_{c}$ and $N_{f}$ are the numbers of colours and the flavours, respectively).

%The dilatons are unstable, they can decay into two photons. 
The dilaton couples with the photons (even before running any SM particles  in the loop) through the trace of the energy-momentum tensor containing the gauge invariant operator of the photon strength tensor. This will cause the dilaton to decay into the photons. It means  the CA acts as a source of the primary (direct) photons (or soft photons in heavy-ion collisions [16]) not produced in decays of light hadrons. The operator associated with the primary photon defines the highest weight of representation of the conformal symmetry and this operator obeys the unitarity condition $d\geq j_{1}+j_{2} +2 - \delta_{j_{1}j_{2},0}$ for the scaling dimension $d$, where $j_{1}$ and $j_{2}$ are the Lorentz spin operators (the "primary" operator means not a derivative of the another operator).
There is an importance of  the primary photons emission for the study of the early Universe and the evolution pattern of the heavy ion collisions, especially in the very early phase. For example, the correlations of direct photons can shed the light to the space-time distribution of the hot matter prior to freeze-out  [17]. However, the investigation of direct photons meets the considerable difficulties mainly due to very small yield of photons emitted directly from the hot dense zone in comparison to the huge background of photons produced due to the standard decay of the hadrons. 
If the PT is situated in some regions accessible to heavy ion collisions experiments one should identified it through some observables. 
%If the CP is situated in some regions accessible to heavy ion collisions experiments one should identified it through some observables to be discovered experimentally. 
 The signature of the PT is the non-monotonous behaviour of observable fluctuation where the latter increases very crucially once the PT approaches. 

We investigate the possible evidence of the DM candidate from the approximate scale symmetry. The DM is the lightest hidden scalar field which is likely a dilaton. 
We suggest the novel approach to the PT and to the approximate scale symmetry breaking with the challenge phenomenology where the primary photons induced by the CA  are in fluctuating regime.
 The rate of the correlation length related to the mass of the dilaton plays an important role as an indicator of the PT achievement if this length grows very sharply. So, the primary photons effects start to play  a significant role when the correlation length grows up intensively in the proximity of the PT or the critical point (CP). 
 The scalar dilaton could be warm and may have the novel feature of BEC into compact massive boson stars [18] leading to direct detection through the observation of primary photons. 
 %In the particle physics experiments it could also be tested if there exist the interactions of the dilaton field $\chi$ with the Higgs boson field operator ${\vert h\vert}^{2}$, $\sim const {\vert h(x)\vert}^{2} \sum_{k=1}^{N} c_{k}\chi_{k}(x)$ %in the sliced-operator form (the tower of hidden dilatons) with $c_{k}$ being the coefficients and $N$ is the number of particles.   
%The definition of observables indicating the location of CP would an instructive challenge.
The paper is organised as follows: in Sec.2, we present the formal model of the dilaton we study. Then, in Sec.3, we summarise the DM scenario with finite temperature and the chemical potential. The condensation and the fluctuations of the scalar particles are presented in Sec.4. In Sec.5, the emission of the primary photons from the BS and the fluctuation rate of the photon production are studied. Finally, we conclude in Sec.6.

\section{The dilaton model}
The dilaton is introduced in the effective field theory by the Lagrangian density (LD) [19]
\begin{equation}
\label{e4}
L = \sum_{i} g_{i}(\zeta)\,O_{i}(x),
\end{equation}
where the local operator $O_{i}(x)$ has the scaling dimension $d_{i}$. The LD (\ref{e4}) is invariant under dilatation transformations $x^{\mu}\rightarrow e^{\omega} x^{\mu}$ with $O_{i} (x)\rightarrow e^{\omega d_{i}} O_{i}(e^{\omega} x)$ if $g_{i}(\zeta)$ is replaced by $g_{i}(\zeta)\rightarrow g_{i} [ (\zeta\,e^{\sigma})]\,e^{\sigma(4-d_{i})}$, where $\sigma(x)$ as the conformal compensator introduces a flat direction; $\omega$ is a $c$-number.
% transforms  according to $\chi(x)\rightarrow e^{\omega}\chi (e^{\omega} x)$.  The order parameter $%\bar f_{\chi}= \langle\chi\rangle$ for scale symmetry breaking is dictated by the dynamics of a relevant %strong sector. 
Obviously, the theory is invariant under scaling transformation if $d_{i} =4$.
The dilaton field is parameterised as $\sigma (x) \rightarrow \chi (x) = f_{\chi}\, e^{\sigma (x)}$, which transforms non-linearly under $\sigma (x) \rightarrow \sigma( x e^{\omega}) + \omega$. The $f_{\chi} = \langle \chi\rangle $ is the vacuum expectation value  of the order parameter $\chi$ for the scale symmetry breaking determined by the dynamics of the underlying strong sector. 
%The dimensional parameters $f_{\pi}$ and $f_{\chi}$  enter in the form of c-number ratio $(f_{\pi}/f_{\chi}){^2}$.
The constant $f_{\chi}$ is defined from 
%dilatation current $S_{\mu}$ acting on the vacuum  $\vert 0\rangle$ defines $\chi$: 
$$\langle 0 \vert S^{\mu}\vert \chi (p)\rangle = i\,p^{\mu}\,f_{\chi} , \,\,\,
\partial_{\mu} \langle 0\vert S^{\mu} (x)\vert \chi (p)\rangle  = -f_{\chi}\,m_{\chi}^{2}\, e^{-i\,p\,x}, $$
where 
$$\langle 0\vert \theta^{\mu\nu} (x)\vert \chi (p)\rangle = f_{\chi} (p^{\mu} p^{\nu} - g^{\mu\nu} p^{2} ) e^{-i\,p\,x}, \,\,\, p^{2} = m_{\chi}^{2}. $$
%and $\langle 0\vert$ is the vacuum state corresponding to spontaneously broken dilatation (and chiral) symmetry.
%$\langle 0 \vert \theta^{\mu}_{\mu}\vert \chi\rangle = m^{2}_{\chi}\,f_{\chi}$.
%with its mass $m_{\chi}$ of the order of QCD scale $\Lambda$,  $\langle 0 \vert \theta^{\mu}_{\mu}\vert \chi\rangle = m^{2}_{\chi}\,f_{\chi}$. 
Here, $S^{\mu} = \theta^{\mu\nu} x_{\nu}$ is the dilatation current, $\theta_{\mu\nu}$ is the energy-momentum tensor,  $m_{\chi}$  is the mass of the dilaton, $p_{\mu}$ is the momentum conjugate to $x_{\mu}$ and $\vert 0\rangle $  is the vacuum state corresponding to spontaneously broken chiral and dilatation symmetries.
%$f_{\chi}$ is the decay constant of a dilaton. 
% otherwise the scale symmetry can be locally restored by introducing a dilaton.
We suppose that at the scale $\geq\Lambda$  the dilaton is formed as the glueball $\chi = O^{++}$, with the mass $m_{\chi} \sim O(\Lambda)$. 
%These dilaton fields interact directly each other in the effective theory, which is a consequence of the microscopic hidden gluon interactions.
The $\chi$ representing the gluon composition $\langle G_{\mu\nu}G^{\mu\nu}\rangle$ responsible for the QCD trace anomaly [20] 
 may be understood as the string ring solution  so that the ends of the string meet each other to form a circle with some finite radius. 
 There is a direct connection between the trace of the energy-momentum tensor $\theta_{\mu}^{\mu}$ and the gluon field-strength tensor $G_{\mu\nu}^{a}$, $\theta_{\mu}^{\mu} = [\beta(\alpha_{s})/(4\,\alpha_{s})] G_{\mu\nu}^{a}\, G^{a\,\mu\nu}$, where $\alpha_{s}  = \alpha_{s} (M)$ is the renormalised gauge coupling defined at the scale $M$; $\beta(\alpha_{s})$ is the renormalisation group $\beta$-function. 
 The glueball becomes massive via the non-vanishing gluon condensation $\langle \chi\rangle \neq 0$ [21]. 
The characteristic feature of the PT is the very sharp increasing of the correlation length $\xi$. The latter describes the fluctuations of the order parameter $\chi$ and acts as a regulator in the IR with  $\xi \sim m_{\chi}^{-1}$. 
If we assume that  $m_{\chi}$ is the continuous function of the temperature $T$, then there might be a phase transition at the critical temperature $T_{c}$ with $m_{\chi} (T_{c}) = 0$. 
The correlation length $\xi$ is not measured directly, however, it influences the fluctuations of the observables, e.g., the primary photons to which the critical mode couples. 
Actually, in the vicinity of the PT, $\xi$ is much larger than that of the size of the particle interacting region at early times.
The scalar $\chi$ may be  a mediator between the conformal sector and the SM. 
However, at high enough temperatures  the role of the mediator is smeared, the scalar field becomes massless in the limit in which the conformal symmetry is recovered, $\theta ^{\mu}_{\mu} = 0$, and the PT is approached.

\section{The DM at finite temperature and chemical potential}
We consider the model containing the dilatons as almost ideal weakly interacting gas (e.g., the glueball gas) at finite temperature. 
%It may correspond to the state of almost quark-gluon plasma in terms of ideal liquid with minimal viscosity.
In case of statistical equilibrium at  temperature $T=\beta^{-1}$ the partition function for $N$ quantum states is 
\begin{equation}
\label{e5}
 Z_{N} = Sp\, e^{-H \beta}, 
\end{equation}
where $H$ is the Hamiltonian 
$H = \sum_{1\leq j\leq N} H(j)$
and $\beta$ in (\ref{e5}) differs from those of the  $\beta$ - function.
For the system of the dilaton functions $\chi_{f}(x)$ which are regular functions in the $f$ - representation, one has the equation $H(j)\chi_{f}(x_{j}) = F(f)\chi_{f}(x_{j})$, where 
\begin{equation}
\label{e6}
H = \sum_{f}\,F(f)\,b^{+}_{f}\,b_{f} = \sum_{f}\,F(f)\,n_{f}
\end{equation}
in terms of the creation and the annihilation operators, $b^{+}_{f}$ and $b_{f}$, respectively; $n_{f}$ is the occupation  number. Here, $F(f) = E(f) - \mu\,Q (f)$ with $E(f)$ being the energy, $\mu$ is the chemical potential of the dilaton associated with  
% $Q(f)$ is the conserved charge with an average density
%$$\langle q\rangle = \frac{1}{\Omega}\,\langle Q\rangle = \frac{1}{\Omega}\,\frac{1}{\beta}\,\frac{\partial}{\partial\mu}\,\ln Z_{N}, $$
%where $\Omega$ is the volume of the system. 
%One has  the open system with a thermal contact and a particle interaction with a reservoir, 
$Q$, which is the operator $N_{f}$ of particles of the type $f$ with the mean value $Tr \{\rho\,N_{f}\} = \hat n_{f}\,\Omega$. The $\rho$ is the statistical operator,  $\hat n_{f}$ is the density of particles of the type $f$ in the volume $\Omega$. 
%Note that in the model with a complex scalar dilaton field the chemical potential $\mu$ looks like the temporal component of a gauge field.
%The interactions between the dilatons should lead to the thermal equilibrium, and in the case of large $n_{f}$ - to the formation of the BS. The gravitational interaction is not an exception [22,23].
In principle, the operators $b_{f}$ in (\ref{e6}) can be distorted by the random quantum fluctuations (e.g., by gluons) in terms of the operator $r_{f}$, $b_{f}\rightarrow b_{f} = a_{f} + r_{f}$, where $a_{f}$ is the bare  operator.  
%In terms of $F(f)$ and $n_{f}$ 
We consider the function $Z_{N}$ (\ref{e5}) in the form 
\begin{equation}
\label{e7}
Z_{N} = \sum_{... n_{f} ... } e^{-\beta \sum_{f} F({f}) n_{f}}.
%Z_{N} = \sum_{\substack {... n_{f} ... \\  \sum_{n_{f} = N}} e^{-\beta\,\sum_{f} F({f})\,n_{f}}.
\end{equation}
%where $\sum_{f} n_{f} = N$.
Since all the  operators $... n_{f} ..$ commute to each other, they may be indentified as the observables. 
The calculation of (\ref{e7}) meets the difficulties because of the condition $\sum_{f} n_{f} = N$  in the limit $N\rightarrow\infty$ in the final stage calculations. This is important because of the particle decays: the dilatons (glueballs) are unstable,  they  may decay into two (primary) photons  which, in case of the photons showers, are registered as a signal that the proximity of PT is approached. The interactions between the dilatons should lead to the thermal equilibrium, and in the case of large $n_{f}$ - to the formation of the BS. The gravitational interaction is not an exception [22,23].

%The phase transitions are characterised through the singularities (discontinuities) in the dependence of various observables on the parameters, e.g., temperature, chemical potential etc. 
The PT manifests itself through the critical chemical potential $\mu_{c}$ and the critical temperature $T_{c}$. Let us consider the probability to formation and evolution of the BS through the following power series  
\begin{equation}
\label{e8}
P(\bar\mu) = \sum_{N =1}^{\infty} Z_{N}\,\bar\mu^{N},\,\,\, \bar\mu = \mu/\mu_{c}.
\end{equation}
%which is the function on $\mu$ and $\beta$,  
%where $\bar\mu = \mu/\mu_{c}$.
% $\mu_{c}$ is the chemical potential in the vicinity of CP. 
Having in mind (\ref{e7}), one has
\begin{equation}
\label{e9}
P(\bar\mu) = \sum_{... n_{f} ...} e^{-\beta\sum_{f} F(f)n_{f}}\,\bar\mu^{\sum_{f} n_{f}} = \prod_{f}\frac{1}{1-\bar\mu e^{-F(f)\beta}}.
%P(\bar\mu) = \sum_{... n_{f} ...} e^{-\beta\sum_{f} F(f)n_{f}}\,\bar\mu^{\sum_{f} n_{f}} = \prod_{f}\left [\sum_{0\leq n < \infty} e^{-\beta F(f)n}{\bar\mu}^{n}\right ] = \prod_{f}\frac{1}{1-\bar\mu e^{-F(f)\beta}}.
\end{equation}
We consider for simplicity that $F(f) \geq 0$ in (\ref{e9}).
Actually,  the convergence radius $R$ of the series (\ref{e8}) will not be less than 1. In the vicinity of PT ($\bar\mu \simeq 1$) one has $\mu_{c} < E(f)/Q(f)$. 
%for $E = {\vert\vec p\vert}^{2}/(2\,m_{\chi})$, where  $\vert\vec p\vert$ is the momentum of $\chi$.

Let us consider (\ref{e8})  in the form
\begin{equation}
\label{e10}
\frac{P(\bar\mu)}{\bar\mu^{N}} = \sum_{N^{\prime} = 0}^{\infty} \frac{Z_{N^{\prime}}\,\bar\mu^{N^{\prime}}}{\bar\mu^{N}}
\end{equation}
on the real axis $0 < \bar\mu <R$. Because of positive $Z_{N^{\prime}}$, the function (\ref{e10}) has the only one minimum on $(0,R)$ 
%\begin{equation}
%\label{e11}
$$\frac{d^{2}\left [P(\bar\mu)\bar\mu^{-N}\right ]} {d\bar\mu^{2}}= \sum_{N^{\prime} = 0}^{\infty} (N^{\prime} - N)(N^{\prime} - N -1)\,Z_{N^{\prime}}\bar\mu^{N^{\prime} - N -2} >0. $$
%\end{equation}
The function (\ref{e10}) tends to infinity when $\bar\mu\rightarrow 0$ and when $\bar\mu\rightarrow R$. In the interval $(0,R)$, there is a point $\bar\mu = \bar\mu_{0}$ at which  (\ref{e10}) has a single minimum, i.e. 
\begin{equation}
\label{e12}
  \frac{d}{d\bar\mu}\left [P(\bar\mu)\,\bar\mu^{-N}\right ]_{\vert_{\bar\mu =\bar\mu_{0}}} = \sum_{N^{\prime} = 0}^{\infty} Z_{N^{\prime}}\, (N^{\prime} - N)\,\bar\mu^{N^{\prime} - N -1}_{\vert_{\bar\mu = \bar\mu_{0}}} = 0. 
\end{equation}
If one goes alone the vertical axis, the ratio (\ref{e10}) has a maximum at $\bar\mu_{0}$. As long as $\bar\mu < \bar\mu_{0}$, no state with $Q \neq 0$ can compete with the vacuum state ($ E = 0$, $Q = 0$) for the role of the ground state.  In case when $\bar\mu > \bar\mu_{0}$, the point $\bar\mu_{0}$ is the ground state at given $\mu$. 

In the physical phase space, the spectrum of the "quasi-momenta" $f$ is almost continuous, and there will be an exact continuous spectrum in the limit $\Omega\rightarrow\infty$. The number $\Delta N$ of different $\Delta f$ in the volume 
$\Omega$ is $(\Delta N/\Delta f) = const\cdot \Omega$. Taking into account (\ref{e9}) and
%Having in mind that (see (\ref{e9}))
%\begin{equation}
%\label{e13}
$$ P(\bar\mu) = \exp\left\{ -\sum_{f}\ln \left [ 1-\bar\mu\,e^{-F(f)\,\beta}\right ]\right\}, $$
%\end{equation}
one can find the asymptotic equality for large $N$
$$ \sum_{f}\ln \left [ 1-\bar\mu\,e^{-F(f)\,\beta}\right ] = N\,\Phi (\bar\mu),$$
where
%$$ \Phi (\bar\mu) = const \cdot v\int \ln \left [ 1-\bar\mu\,e^{-F(f)\,\beta}\right ] df, \,\,\, v = \frac{\Omega}{N}.$$
$\Phi (\bar\mu) = const \cdot v\,\beta\,K_{\chi}(\bar\mu)$, $ v = \Omega/{N}$. 
The thermochemical potential  $K_{\chi}(\bar\mu)$ of the dilaton $\chi$
\begin{equation}
\label{e14}
K_{\chi}(\bar\mu) = \beta^{-1}\,\int \ln \left [ 1-\bar\mu\,e^{-F(f)\,\beta}\right ] df
\end{equation}
gives the contribution to the thermodynamic potential 
\begin{equation}
\label{e15}
 K = K_{\chi} + V_{\chi} + \lambda \left ( \frac{f_{\chi}}{2}\right )^{4}.
\end{equation}
Here, $V_{\chi}$ is the potential term in the LD of the dilaton 
$L_{\chi} = (1/{2})\partial_{\mu}\chi\partial^{\mu}\chi - V_{\chi},$
$$V_{\chi} = \frac{\lambda}{4}\chi^{4}\,\left ( \ln\frac{\chi}{f_{\chi}} -\frac{1}{4}\right ),$$
where $\lambda = m^{2}_{\chi}/f^{2}_{\chi}$.
The term $\lambda (f_{\chi}/2)^{4}$ in (\ref{e15}) is added so that $K = 0$ at $T = 0$ and $\chi = \langle\chi\rangle = f_{\chi}$. 
%In the vicinity of the CP, the free gluons are disfavored as appropriate degrees of freedom in the phase with confinement of quarks. The loops containing the heavy quarks with masses $m_{h}$ can generate radiative corrections to %$m_{\chi}^{2}$ of magnitude $\delta m_{\chi}^{2} \sim m_{h}^{2}\,M_{UV}^{2}/(4\,\pi\,f_{\chi})^{2}$. The latter do not influence at CP ($m_{h}\rightarrow 0)$.

%The thermodynamic potential (\ref{e15}) does account for dilaton (glueball) and gluon degrees of freedom:
In terms of the glueball and the gluon degrees of freedom (d.o.f.), the potential (\ref{e15}) looks like
\begin{equation}
\label{e166}
K = \theta (\beta - \beta_{c})\,K_{\chi} (\bar\mu) + \theta (\beta_{c} - \beta)\,K_{gH} ,
\end{equation}
where $K_{gH} = K_{g} + K_{H}$  is an effective gluon thermodynamic potential with the quasi-gluon energy $E_{g} = \sqrt {{\vert \vec p\vert}^2 + m^{2}_{g}}$. The effective gluon mass $m_{g}$ is introduced for the phenomenological reason. $K_{H}$ is the model-dependent 
Haar measure contribution (see for details [24]). 
The potential $K_{g}$ is  the model-dependent function 
\begin{equation}
\label{e16}
% K_{g} = \beta^{-1}\,\int \ln \left [ 1- e^{-E_{g}(f)\,\beta}\right ] df.
K_{g} \simeq \frac{m^{2}_{g}}{\pi^{2}\,\beta^{2}}\,\sum_{n=1}^{\infty} \frac{C_{n}}{n}\, K_{2}(n \beta m_{g}),
\end{equation}
where the temperature-dependent  gluon mass $m_{g} = m_{g}(\beta) = g(\beta)/\beta$, $g(\beta)$ is the effective gauge coupling; the colour coefficients $C_{n}$ are given in [24]; $K_{2}(x)$ is the Bessel function. 
The quark contribution to the potential (\ref{e166}) at $T < T_{c}$ has the negative sign compared to that of the gluon part. The physical result can restrict the number of quark flavours. At zero chemical potential the combined potential $K (T < T_{c}) $ will be positive up to 3 light quark flavours if the "constituent" mass of the "quasi-gluon", $\sim 0.5\, m_{\chi}\simeq$ 0.85 GeV [25], and the "constituent"  mass of the quark, $\sim$ 0.3 GeV, are used.     
The second term in (\ref{e166}) yields the first-order PT at the critical point as found in the SU(3) lattice calculations [26,27]. The potential (\ref{e166}) indicates that the gluons are forbidden below the critical temperature as the coloured degrees of freedom. 
Both forms (\ref{e14}) and (\ref{e16}) match  each other at the PT.
For further study where the PT is the well-defined singularity with $T$ and  $\mu$, we use the function
%To scan the parameters relevant to the PT  we use the function
\begin{equation}
\label{e17}
 P(\bar\mu)\,\bar\mu^{-N} = \left [\bar\mu^{-1}\,e^{-\Phi (\bar\mu)}\right ]^{N}.
\end{equation}
It does not concern the phase diagram to scan the critical point  (in QCD) where the position of the latter is not clear from the theoretical point of view.
In our case, the PT is a well-defined singularity with $T$ and  $\mu$. 
%To get closer to the PT, one has
In order to find the PT we need to calculate (\ref{e17}) where the relevant singularity will be visible.
% corresponding to the end of the (first order) transition line.
For this end, let us consider the circle $C$ with the radius $r=\bar\mu_{0}$ with the origin being at zero. The partition function is 
\begin{equation}
\label{e18}
 Z_{N} = \frac{1}{2\,\pi\,i}\int_{C} \frac{P(\bar\mu)}{\bar\mu^{N+1}}\,d\bar\mu \rightarrow \frac{1}{2\,\pi}\int_{-\pi}^{+\pi} \frac{P(r\,e^{i\,\varphi})}{r^{N}\,e^{i\,N\,\varphi}}\,d\varphi.
\end{equation}
%There is a maximum of the function under an integration in (\ref{e11}) at $\varphi = 0$ because of %power $N$ in (\ref{e10}). 
Since $r\neq 0$ the maximum of the function under the integration in (\ref{e18}) is expected at $\varphi = 0$ taking into account the number of particles $N$ in the exponential function in (\ref{e17}). Hence, the calculation of (\ref{e17}) depends on  the behaviour of the function under the integration in (\ref{e18}) at $\varphi = 0$.  
Taking into account the minimum condition ({\ref{e12}), we have
$$\frac{\partial}{\partial\varphi}\left [ \frac{P(r\,e^{i\,\varphi})}{r^{N}\,e^{i\,N\,\varphi}}
\right ]_{\vert_{\varphi = 0}} = 0 .$$
%Hence,
%$$ \left [\frac{\partial}{\partial\varphi}\,\Phi (r\,e^{i\,\varphi}) + i\,\varphi\,\ln r\right ]_{\vert_{\varphi = %0}} = 0.$$
%In the vicinity of $\varphi = 0$ one can find 
%\begin{equation}
 %\label{e12}
 %\left [\frac{e^{-\Phi (\bar\mu)}}{\bar\mu}\right ]^{N}\rightarrow \left [\frac{e^{-\Phi (r)}}{r}\right ]^{N}%%\,e^{-N\,\alpha\,\varphi^{2}},
%\end{equation}
%where $\alpha = - \Phi^{\prime\prime} (r) - \Phi^{\prime} (r) > 0 $ comes from the expansion 
%$ \Phi (r\,e^{i\,\varphi}) + i\,\varphi\,\ln r = \Phi (r) + \alpha\,\varphi^{2} + ...$ Using (\ref{e12}) one can %find 
%The asymptotic form of partition function $Z_{N}$ (\ref{e11}) 
%\begin{equation}
 %\label{e13}
 %Z_{N} = \frac{1}{2\,\pi}\int_{-\infty}^{+\infty} \frac{e^{-N\,\Phi (r)}}{r^{N}}\,e^{-N\,\alpha\,\varphi^{2}} d%\varphi = \frac{P(r)}{2\,r^{N}\,\sqrt{\pi\,\,N\,\alpha}}.
%\end{equation}
The asymptotic form of the partition function is
%\begin{equation}
 %\label{e19}
 $$\ln Z_{N}\simeq -\sum_{f}\,\ln \left [1-\bar\mu_{0}\,e^{-F(f)\beta}\right ] - N\,\ln\bar\mu_{0} - \ln 2\sqrt {\kappa\,\pi\, N},$$
%\end{equation}
where $\kappa = - \Phi^{\prime} (\bar\mu_{0}) - \Phi^{\prime\prime} (\bar\mu_{0})$. Considering Eq. (\ref{e12}) at $\bar\mu = \bar\mu_{0}$, one can  find 
%where $\bar\mu_{0}$ is defined from the Eq.
%where the term $-\ln (2\,\sqrt{\pi\,\alpha\,N})$ is neglected compared to $N\ln\mu_{0}$ in (\ref{e14}). 
%Using (\ref{e14}) in (\ref{e9}) we get the Eq. to define  $\mu_{0}$ 
%($\sum_{f} \bar n_{f} = N$)
\begin{equation}
 \label{e20}
 %\bar n_{f} = \frac{1}{\mu_{0}^{-1}\,e^{F(f)\beta} -1}
%\mu_{0} = \frac{\bar n_{f}}{1 + \bar n_{f}} \,\exp[{{F(f)\beta}}],
\sum_{f} \bar n_{f} = \sum_{f} \frac{1}{\bar\mu_{0}^{-1}\, e^{F(f)\beta} -1} = N
\end{equation}
and hence, $\bar\mu_{0}$ can be extracted from (\ref{e20}). One can calculate the sum of the quantum states up to the singular point defined by the relation between $\bar\mu_{0}$ and $F\beta$, and large $N$.
We assume the large number $N$ in (\ref{e20}) which is correct if the dilatons are light. This is important in the sense to the proposal of condensed dark matter bosons in the early Universe or at the early stage after the heavy ion collisions. In some sense, it corresponds to  Bose star formation as the compact object of BEC bounded by self-gravity [28].

\section{The condensation and the fluctuations}
The hidden scalar particles could have the correct relic density and be non-relativistic enough.
As the temperature lowers below $\sim \Lambda$, the dilatons can act as a classical form of the DM in the late Universe.  
In this case, the dilaton becomes the warm DM with the density $\sim exp (-\Lambda\,\beta)$ which follows down very rapidly. 
%The dilatons are unstable and they decay to the primary (direct) photons or/and to the dark photons. 
%There are almost negligible interactions between the dilatons and their annihilations weakly influence the final number of DM particles.   
Consider the model where the dilatons are produced in the volume $\Omega$ as a cube with the side of the length $L = \Omega^{1/3}$. We assume the wave function of the dilaton in the form 
$ \phi_{p} (q) = \Omega^{-1/2}\,e^{i\,q\,p}$, where $p^{\alpha} = (2\,\pi/L)\,n^{\alpha}$, $\alpha = 1,2,3$; $n^{\alpha} = 0, \pm 1, ...$ and the energy is $E_{p} = {\vert p\vert}^{2}/(2\,m_{\chi})$.
%In the limits $N\rightarrow\infty$,  $\Omega\rightarrow\infty$ and for $v = const$ 
We consider two cases:
%there are two possibilities one has for $\mu_{0}$ defined by (\ref{e15}): 
the high temperature case A), where $\bar\mu_{0}\,  e^{\mu\,Q\beta} < 1$, and
the low temperature case B), where $\bar\mu_{0}\, e^{\mu\,Q\beta} \sim 1$.

In the case A), the function $\bar n_{f}$  is regular on $f$, and the sum $\sum_{f} \bar n_{f}$ is replaced by the integral
%where the spectrum of $f$ is continuous at $\Omega\rightarrow\infty$:
$$\frac{1}{v} = \frac{1}{\Omega}\,\sum_{f}\bar n_{f}\rightarrow \frac{1}{(2\,\pi)^{3}}\int\bar n(f)\,d^{3} f. $$ Using (\ref{e20}) we arrive at the equality:
\begin{equation}
 \label{e210}
 \int_{0}^{\infty}\frac{x^{2}\,dx}{\bar\mu_{0}^{-1}\,e^{-\mu\,Q\beta}\,e^{x^{2}} -1} = \frac{2\,\pi^{2}}{v}\,{\left (\frac{\beta}{2\,m_{\chi}}\right )}^{3/2}. 
\end{equation}
%under the condition 
%\begin{equation}
 %\label{e22}
%\mu\,Q + T\,\ln\mu_{0} \leq 0. 
%\end{equation}
In the case A), the integral in the l.h.s. of (\ref{e210}) increases if $\bar\mu_{0}\,  e^{\mu\,Q\beta}$ increases as well.
%\begin{equation}
%\label{e201}
%T\rightarrow \mu\,Q\,\ln^{-1} (\bar\mu^{-1}_{0}).
%\end{equation}
  The Eq. (\ref{e210})  has the solution in terms of $\mu Q$ and $\bar\mu_{0}$ when 
\begin{equation}
 \label{e211}
 \frac{2\,\pi^{2}}{v}\,{\left (\frac{\beta}{2\,m_{\chi}}\right )}^{3/2} < \frac{\sqrt \pi}{4}\,B,
 \end{equation}
 where the r.h.s. of inequality (\ref{e211}) is the result of calculation of the integral in the l.h.s. of (\ref{e210}) under the condition $T\rightarrow \mu\,Q\,\ln^{-1} (\bar\mu^{-1}_{0})$, 
 and $B = 2,612...$ is the Riemann's zeta-function, $\zeta (3/2)$. 
%The integral in l.h.s. of (\ref{e21}) increases if $\mu\,Q + T \ln\bar\mu_{0} \rightarrow 0$, that will allow one to find an inequality
%\begin{equation}
% \label{e22}
%$$\frac{2\,\pi^{2}}{v}\,{\left (\frac{\beta}{2\,m_{\chi}}\right )}^{3/2}  < \int_{0}^{\infty}\frac{x^{2}\,dx}{e^{x^{2}} -1} . $$
%\end{equation}
The case A) is realised  when  the temperature $T$ exceeds the critical one, $ T >T_{c}$, where the critical temperature is defined by the dilaton mass and the inverse particle density $v = \Omega/N$ 
%\begin{equation}
 %\label{e23}
$$T_{c} = \frac{2\,\pi}{m_{\chi}} \left (v\,B \right )^{-2/3},  \,\,\,\,m_{\chi} \neq 0.$$
%\end{equation}
%In the plane $(T - \mu)$ the ratio for CP $(\mu\,Q/T)_{c}$  is defined by $- \ln\mu_{0}$,
%CP is defined by the ground state $\bar\mu_{0}$ and depends on the number of particles $N$
One can easily find that the correlation length $\xi$  is defined by $\mu$ and has the dependence of $N$ (through $v$). The singular behaviour of $\xi$ is governed by the ground state chemical potential $\bar\mu_{0}$:
%\begin{equation}
 %\label{e199}
%\left (\frac{\mu\,Q}{T}\right )_{c} = \ln\left (\frac{1}{\mu_{0}}\right ).
%\mu_{c}\,Q = - \ln(\bar\mu_{0})\,\frac{1}{2m_{\chi}}\,\left (\frac{2\,\pi^{2}}{v\,B}\right)^{2/3}.
$$\xi \sim \frac{\mu\,Q}{2\,\pi}  \left (v\,B\right )^{2/3}\,\ln^{-1}\left (\frac{1}{\bar\mu_{0}}\right ). $$
%\end{equation}
%Actually, $\xi\rightarrow\infty$ at $\bar\mu_{0}\rightarrow 1$ that means the vicinity at the CP ($\bar\mu_{0} = \bar\mu =1$).
%If $T\rightarrow T_{c}$ 
The fluctuations of the dilaton mass become more correlated, and the length scale increases toward infinity when  $\bar\mu_{0}\rightarrow  1$. The fluctuations at $T << T_{c}$ have short correlation length ($ \bar\mu_{0} < 1$).
%at $\mu_{0}\rightarrow 1$ from below the CP is shifted to the left in the plane $(\mu - T)$. %otherwise the temperature is doing down when $\mu_{0}\rightarrow 0$. In case of 1-particle state, %where $\mu_{0} = [\bar n_{f}/(1 + \bar n_{f})]e^{F(f)\beta}$, the case A) is nothing other but the high %temperature one:
%$$ T > E(f)\,\ln^{-1} \left (\frac{1 + \bar n_{f}}{\bar n_{f}}\right ),$$
% and the critical condition (\ref{e20}) transforms into 
%$${\left (\frac{p^{2}}{T}\right )}_{c} = 2\,m\,\ln \left (\frac{1 + \bar n_{f}}{\bar n_{f}}\right ),$$
%where the momentum in the critical mode is 
%$$\vert p_{c}\vert = \left (\frac{2\,\pi^{2}}{B\,v}\right )^{1/3}\,\ln^{1/2} \left (\frac{1 + \bar n_{f}}{\bar %n_{f}}\right ).$$
%The case B) corresponds to thermal particle system with $T\sim E(f)/\ln\left (\frac{1 + \bar n_{f}}{\bar %n_{f}}\right ).$
%To estimate the fluctuations of particles we consider the volume $V$ which is less than $\Omega$. %The fluctuation of particle density is 
%\begin{equation}
% \label{e21}
%\langle {(n_{V} - \langle n_{V}\rangle )}^{2}\rangle = \langle n_{V}\rangle 
%\left [ 1 + \frac{\sqrt {2}\,v}{\pi^{2}}\, (m\,T)^{3/2}\int_{0}^{\infty} \frac{ x^{2}\,dx}{(\mu_{0}^{-1}\,e^{-\mu%\,Q\,\beta}\,e^{x^{2}} -1)^{2}} \right ],
%\end{equation}
%where $\langle n_{V} \rangle = V/\Omega$. The sharp increasing of particle fluctutaion is expected %when 
%$\mu Q$ tends to $T\ln(1/\mu_{0})$, and the maximal fluctuations will be at CP (\ref{e20}).
%For finite and small $v $ 
There is no PT caused by developing of the  correlation length $\xi$ if $v$ is finite and small enough. Under the thermal influence the non-monotonic behaviour of $\xi$ is assumed to be as an indicator of the PT 
(see also  [29] and [30] for the case of the QCD phase diagram with the CP).

In the case B), our interest is in small $\vert p\vert \leq\delta$ (maximal $N$), where the CP is approached. Here,
%\begin{equation}
 %\label{e25}3
$$\frac{1}{\Omega}\sum_{\vert p\vert \leq\delta} \bar n_{p} = \frac{1}{v} - \frac{1}{\Omega}\sum_{\vert p\vert \geq\delta} \bar n_{p}, $$
%\end{equation}
where 
%\begin{equation}
 %\label{e26}
$$\lim_{\delta\rightarrow 0,\,\,N\rightarrow\infty} \frac{1}{\Omega}\sum_{\vert p\vert \leq\delta} \bar n_{p} = \frac{1}{v} - \frac{1}{(2\,\pi)^{3}}\int_{\vert p\vert \geq\delta} \frac{d^{3} p}{e^{E_{p}\beta} -1} 
= \frac{1}{v} \left [1-{\left (\frac{\beta_{c}}{\beta}\right  )}^{3/2} \right ]. $$
%\end{equation}
Hence, the case B) takes place for $T < T_{c}$, where the only part of the total number of scalar particles proportional to $\sim (\beta_{c}/\beta)^{3/2}$ is distributed in the BS with all the spectrum of momenta. The rest one $\sim [ 1- (\beta_{c}/\beta)^{3/2}]$ is the scalar condensate. As the result, in the case of high temperatures, the condensates stay close to almost zero, while at low temperatures the condensate obtains a large value. 

Now, one can connect the results for the fluctuations of $\chi$ to the fluctuations of the observables. For this end, we suppose that the particles are in the local volume $V$ of the compact BS which is  less 
then the volume $\Omega$ of the physical phase space. The number of particles $n_{V}$ in $V$ is $\sum_{1\leq j\leq N} \hat n_{V} (q_{j})$, where $\hat n_{V} (q) = 1$ if $ q\in V$, and $\hat n_{V}( q) = 0$ otherwise.  The volume $V$ is defined by the geometry of the BS. 
The event-by-event fluctuation of the particle density $\langle {(n_{V} - \langle n_{V}\rangle )}^{2}\rangle $  is 
% increases sharply when $T/\mu \rightarrow Q/\ln(1/\bar\mu_{0})$ 
\begin{equation}
 \label{e27}
%\bar n_{V}^{2} = 
%\langle n_{V}\rangle 
%\left [ 1 + \frac{\sqrt {2} v}{\pi^{2}} \left (\frac{T}{\xi}\right )^{3/2}\int_{0}^{\infty} \frac{ x^{2}dx}{(\bar\mu_{0}^{-1}e^{-\mu Q \beta} e^{x^{2}} -1)^{2}} \right ], 
\frac{\langle {(n_{V} - \langle n_{V}\rangle )}^{2}\rangle}{ \langle n_{V}\rangle} -1 = \frac{\sqrt{2}\,v}{\pi^{2}} \left (m_{\chi}\,T\right )^{3/2} \int_{0}^{\infty} \frac{ x^{2}dx}{(\bar\mu_{0}^{-1}e^{-\mu Q \beta} e^{x^{2}} -1)^{2}} , 
\end{equation}
where $\langle n_{V} \rangle = V/\Omega$. Obviously, (\ref{e27}) increases sharply when the temperature $T \rightarrow \mu\,Q/\ln(1/\bar\mu_{0})$. 
%The non-monotonous behaviour of the particle density fluctuation is evident 
%when  $\mu Q$ tends to $T\ln(1/\mu_{0})$, and the maximal fluctuations will be 
When the PT approaches, the fluctuation (\ref{e27}) at the CP is
%disappears at the critical temperature with the result where no free parameters one has
%where there is no the dependence of particle multiplicity:
\begin{equation}
 \label{e28}
% \frac{\langle {(n_{V} - \langle n_{V}\rangle )}^{2}\rangle}{ \langle n_{V}\rangle} -1 \simeq \frac{4}{\sqrt{\pi}\,B} \int_{0}^{\infty} \frac{ x^{2}\,dx}{(e^{x^{2}} -1)^{2}}. 
\frac{\langle {(n_{V} - \langle n_{V}\rangle )}^{2}\rangle}{ \langle n_{V}\rangle} -1 \simeq \frac{4}{\sqrt{\pi}\,B} \int_{0}^{\infty} \frac{ x^{2}\,dx}{(z_{c}\,e^{x^{2}} -1)^{2}},
\end{equation}
where $z_{c} = \bar\mu_{0}^{-1} e ^{-a_{c}}$, $a_{c} \simeq \mu_{c}\,Q\,\Lambda\, (vB)^{2/3}/(2\,\pi)$. One can expect the non-monotonous rising of the fluctuation (\ref{e27}) when  $\bar\mu_{0}\rightarrow 1$ with $v\rightarrow 0$ at infinite number $N$ of particles. There are no free parameters  in (\ref{e28}) when the CP is approached.

\section{The primary  photons from the BS}

In the exact scale symmetry, $\chi$ couples with the SM particles via the trace of $\theta_{\mu\nu}$, $L = (\chi/f_{\chi}) ( \theta^{\mu}_{\mu_{tree}} +  \theta^{\mu}_{\mu_{anom}} )$,
%\begin{equation}
%\label{e29}
 %L = \frac{\chi}{f_{\chi}} \left ( \theta^{\mu}_{\mu_{tree}} +  \theta^{\mu}_{\mu_{anom}}\right ),
%\end{equation}
where the first term contains
%in (\ref{e29}) 
%(the contributions from heavy quarks and heavy gauge bosons are neglected)
%$$\theta^{\mu}_{\mu_{tree}} = -\sum_{q} [m_{q} + \gamma_{m}(g) ]\bar q q + 2 m^{2}_{W} W^{+}_{\mu} %W^{{-}\,\mu} + m^{2}_{Z} Z^{\mu} Z_{\mu} - \frac{1}{2} m^{2}\chi^{2} +\partial_{\mu}\chi\partial^{\mu}%\chi, $$
$$\theta^{\mu}_{\mu_{tree}} = -\sum_{q} [m_{q} + \gamma_{m}(g) ]\bar q q  - \frac{1}{2} m_{\chi}^{2}\chi^{2} +\partial_{\mu}\chi\partial^{\mu}\chi. $$
 Here, $q$ is the quark field  with the mass $m_{q}$; $\gamma_{m}$ are the corresponding anomalous dimensions, and the contributions from heavy quarks and heavy gauge bosons are neglected. In contrast to the SM, the dilaton couples with massless gauge bosons even before running any SM particles in the loop, through the trace anomaly.
The latter has the following term   for the photons and the gluons:
\begin{equation}
\label{e299}
\theta^{\mu}_{\mu_{anom}} = -\frac{\alpha}{8\,\pi}\, b_{EM}\, F_{\mu\nu}F^{\mu\nu} - 
\frac{\alpha_{s}}{8\,\pi}\sum_{i}\, b_{0_{i}}\, G_{\mu\nu}^{a}G^{{\mu\nu\,a}},
\end{equation}
where $\alpha$ is the fine structure coupling constant, $b_{EM}$ and $b_{0_{i}}$ are the coefficients of the electromagnetic (EM) and the QCD $\beta$-functions, respectively. If the strong and the EM interactions are embedded in the conformal sector, the following relation for light and heavy particles sectors is established above the scale $\Lambda$: 
$\sum_{light} b_{0} = - \sum_{heavy} b_{0}$, where the mass of $\chi$ splits the light and heavy states. The anomaly  term for gluons in (\ref{e299})
$$ \frac{\beta (g)}{2\,g} G_{\mu\nu}^{a}G^{{\mu\nu\,a}} = \frac{\alpha_{s}}{8\,\pi}\,b^{light}_{0}  G_{\mu\nu}^{a}G^{{\mu\nu\,a}}, \,\,\, b^{light}_{0} = -11 +\frac{2}{3}n_{L} $$
is evident, where the only $n_{L}$ particles lighter than $\chi$ are included in  the $\beta$ - function, $\beta (g) = b^{light}_{0}\,g^{3}/(16\, \pi^{2})$, $ g^{2} = 4\pi\alpha_{s}$. For example, for $m_{\chi}\sim O(\Lambda)$ one has $n_{L} = 3$ that indicates about 14 times increase of the dilaton-gluon-gluon coupling strength compared to that of the SM Higgs boson. 
%The $\beta$ - function vanishes above the scale of spontaneous breaking of scale invariance.

The light dilaton operates with the low invariant masses, where two photons are induced effectively by gluon operators. In the low-energy effective theory valid below the conformal scale $ \sim 4 \pi f_{\chi}$ at small transfer-momentum $q$,  
$\langle \gamma\gamma\vert \theta^{\mu}_{\mu} (q)\vert 0\rangle\simeq 0$  [31]
and 
$$\langle \gamma\gamma\vert b^{light}_{0}\,\alpha_{s}\,
 G^{a}_{\mu\nu} G^{\mu\nu\,a}\vert 0\rangle = - \langle \gamma\gamma\vert b_{EM}\,\alpha\, F_{\mu\nu} F^{\mu\nu}\vert 0\rangle, \,\, \vec q = 0.$$
The partial decay width $\chi\rightarrow\gamma\gamma$ is
\begin{equation}
\label{e21}
 \Gamma (\chi\rightarrow\gamma\gamma)\simeq \left (\frac{\alpha\,F_{anom}}{4\pi}\right )^{2}\,\frac{m_{\chi}^{3}}{16\,\pi\,f_{\chi}^{2}},
 \end{equation}
where the only CA contributes through 
$$F_{anom} = -(2 n_{L}/3) (b_{EM}/b^{light}_{0}), \,\,\,b_{EM} = -4\sum_{q:u,d,s} e^{2}_{q} = -8/3, $$ 
$e_{q}$ is the charge of the light quark. When one approaches the PT (the first-order transition), the absolute value of $F_{anom}$ decreases due to increasing on $b^{light}_{0}$ as $n_{L}\rightarrow 0$. 
There are fluctuations of the dilaton field with the finite mass $\sim O(\Lambda)$ which is model-dependent. 
In the proximity to the IRFP, the dilaton mass is $m_{\chi}\simeq \sqrt{1-N_{f}/N^{c}_{f}}\Lambda$ [32], where $N^{c}_{f}$ is the critical value of $N_{f}$ corresponding to $\alpha^{c}_{s}$ at which the chiral symmetry is breaking and the confinement arises. On the other hand, from lattice calculations [33,34], the lightest glueball masses approach a constant at large number $N$ of colours in the hidden $SU(N)$ sector, and can be well defined as $m_{\chi} = (a + b/N^{2})\Lambda$. Using these two parameterisations, one can conclude that $a$ should be less than 1, and $b\sim O(1)$. The self-coupling $\lambda $ in  (\ref{e15}) is very large, $\lambda \simeq 112 $,  if one uses the glueball mass $m_{\chi} = 1.7$ GeV [25] and the vacuum energy density $E = \lambda (f_{\chi}/2)^{4} = 0.6$ GeV$fm^{-3}$ [35]. On the other hand, $\lambda \sim O(1)$ that comes from  $\lambda \simeq (1 - N_{f}/N^{c}_{f})\simeq (a + b /N^{2})^{2}$, if $m_{\chi}\sim O(\Lambda)$ and $f_{\chi} \simeq \Lambda$. 
%When one approaches the PT (the first-order transition) the absolute value of $F_{anom}$ decreases due to increasing on $b^{light}_{0}$ as $n_{L}\rightarrow 0$. 
%The second-order phase transition is characterised by the limits $N_{f}\rightarrow N^{c}_{f}$ and $\Lambda\rightarrow 0$, hence no primary photons should be evident through a detector. 

Because of the strong couplings of the dilatons with gluons and photons one can expect the abundant production of the dilatons (glueballs) due to gluon-gluon fusion and the decays of the dilatons (glueballs) to primary photons. 
%In the early Universe, the scalar glueballs are unstable and it is expected the showers of primary photons including the dark photons [35] due to conformal gluon and electromagnetic anomalies. 
%In the approximate conformal sector (the proximity of the PT) the fluctuation rate of the photons escape is
The measurement of the photons escape in the early stage of the heavy-ion  collisions provides a decisive way to observe and to differentiate the primary photons and the ordinary photons in the decays  of the secondary produced light hadrons, e.g., $\pi^{0}\rightarrow \gamma\gamma$. The trace-anomaly term ({\ref{e299}) contributes for dilatons but not for light hadrons. It is only necessary to count the event numbers of $\gamma\gamma$ in the heavy-ion collisions at different energies. In the confinement stage, the effects of the explicit scale symmetry breaking are dominant, and it is assumed that the confinement triggers the spontaneous breaking the chiral symmetry for light quarks. One can estimate the fluctuation rate of the photon production in the approximate conformal sector (the proximity of the PT) via the rate
\begin{equation}
\label{e209}
%r_{\gamma\gamma} = 1 + \frac{BR (\pi^{0}\rightarrow \gamma\gamma)}{BR (\chi\rightarrow \gamma\gamma)} = 1+ m_{\pi}^{3} \,\left (\frac{6}{F_{anom}}\right )^{2}\,\xi^{3},
r_{\gamma\gamma} = 1+ m_{\pi}^{3} \,\left (\frac{6}{F_{anom}}\right )^{2}\,\xi^{3}.
%\simeq \frac{\Gamma (\pi^{0}\rightarrow \gamma\gamma)}{\Gamma (\chi\rightarrow \gamma\gamma)}, 
\end{equation}
%where 
%$r_{\chi}$ is related with the cross-sections $\sigma$ for production of $\pi^{0}$-mesons and the dilatons $\sigma (\pi^{0}) = \sigma (\chi)\cdot r_{\chi}$. In (\ref{e209}), 
%$BR (P\rightarrow\gamma\gamma)$ is the branching ratio in the process  $ P\rightarrow\gamma\gamma $ ($P: \pi^{0}, \chi$).
At the PT, one can expect the sharp rising of the fluctuation rate $ r_{\gamma\gamma}$ in the  IR ($\alpha^{\star}_{s} > \alpha^{c}_{s}$) relevant to  direct photons 
where at large distances we use the effective d.o.f. in terms of neutral $\pi^{0}$ - mesons with the mass $m_{\pi}$. The abundant escape of the photons will be as $\xi (T\rightarrow T_{c})\rightarrow \infty $ and $N_{f}\rightarrow N_{f}^{c} $. The critical value $N_{f} = N_{f}^{c}$ separates the conformal phase from the one with confinement and massless quark formation in the frame to the chiral gauge theory. The method is independent of the values of the model parameters, where for an order of magnitude one can take $f_{\chi}\simeq \Lambda$ and the pion constant $f_{\pi}\simeq 0.3 \Lambda$.
 The result (\ref{e209}) is consistent with the physical pattern where the dilaton is emerged at the scales $\geq \Lambda$ as well as $\pi^{0}$'s and other light quark bound states.
It is easily to find that at the PT, $r_{\gamma\gamma}\rightarrow\infty$ when the number of light quarks $n_{L}\rightarrow 0$. 
%The latter is the consequence of the very small mass $m_{\chi}$ of the dilaton at $N_{f}\rightarrow N_{f}^{c}$
%as well as  $\alpha^{\star}_{s}\rightarrow \alpha^{c}_{s}$.
%On the other hand, $R_{\chi} = 0$ should be viewed as an UV fixed point of the strong coupling phase %transition.
%The contribution with (\ref{e209}) is counting by the detection of (primary) photons which indicate the region of the PT and the critical point where the escape and detection of the photons are maximal. 
 Thus, one can expect to find the non-monotonous increasing of the photons fluctuations once  it is going away from  UV to IR. The contribution with (\ref{e209}) is counting by the detection of the primary photons which indicate the region of the PT and the CP where the escape and detection of the photons are maximal. 
 The measurement of the photon fluctuations can be used to determine whether the quantum system is in the vicinity of the PT or not.
 
% \section{$SU(N)$ hidden sector in the BS }
 %In the beginning of this section let us note that the scalar dilaton in the form of the glueball is a well-motivated example coming from a pure Yang-Mills hidden sector, which locks up into the bound state of two gluons in the early Universe. The glueball DM may condense into the boson stars and be observed by gravitational  lensing effect.  e 
Finally, let us note that in the paper [7], there was presented the direct point-like coupling of the hidden sector associated with the scalar glueball field $\chi$ in the BS to the photons in the hidden sector of the SU(N) gauge theory with an unspecified  value of $N$
%\begin{equation}
%\label{e22}
 $$ \frac{1}{M_{cut}^{4}} H_{\mu\nu}H^{\mu\nu} F_{\alpha\beta}F^{\alpha\beta} \rightarrow  \frac{N\,m_{\chi}^{3}}{M_{cut}^{4}} \chi\, F_{\alpha\beta}F^{\alpha\beta}, $$
% \end{equation}
 where $H_{\mu\nu}$ and $F_{\alpha\beta}$ are the strength tensors of the hidden gauge field of the group $SU(N)$  and the photon, respectively; $M_{cut}$ is the cutoff scale. The decay rate of $\chi$ into two photons in the point-like (direct) interaction in the star is 
 \begin{equation}
\label{e22}
\Gamma (\chi\rightarrow\gamma\gamma) = \frac{1}{4\pi} m_{\chi}\,N^{2} {\left (\frac{m_{\chi}}{M_{cut}}\right )}^{8},
\end{equation}
where the value $N$ making $\chi$ a self-interacting hidden matter is $N\simeq Max [(0.1 GeV/m_{\chi})^{3/4}, 2]$. The combined result in the conformal anomaly (\ref{e21}) and the direct interaction (\ref{e22}) between the scalar glueball and the photons  gives the strongest constraints on the scale $M_{cut}$ with the DM mass in the MeV's scale.  We find that for  $m_{\chi}\sim O(\Lambda)$ the cutoff $M_{cut}$ is allowed to be as low as the weak scale: $M_{cut} \geq 3.4 $ GeV and $M_{cut} \geq 5.2 $ GeV for $\Lambda = 330$ MeV and  $\Lambda = 500$ MeV, respectively.

%Comparing the results in (\ref{e21}) and in (\ref{e22}), one can find $N\simeq 2$ if the masses of the dilaton $\chi$ and the glueball 
%$\phi$ fields are equal to each other; $f_{\chi} \sim O(\Lambda)$. The estimation for $N$ is done for the chosen parameters $n_{L} = 3$, $\Lambda = 1$ GeV and $M = 10$ GeV. The cosmic ray photon observations constrain  the %hidden glueball decay into photons where  $M$ is allowed to be as low as the weak (or TeV) scale if $m$ is in the range from keV to MeV (see Fig.4 in [29]).

\section{Conclusions}
To conclude, we investigated the possible evidence of the DM candidate  from the approximate conformal symmetry. The DM is the lightest hidden scalar field which is likely the dilaton or the glueball. 
These scalar fields could be warm and may have the novel feature of the BEC into the compact massive boson stars. 
The importance of the BS is actual and evident, because the boson stars hide a scalar part of DM from direct observation. After the formation and becoming large in size, the BS may explode into leptons via decays of the dark photons [36] or may emit  the primary photons which  may explain the conformal anomaly in EW and EM hidden sectors. These could contribute to new sources of cosmic rays.
%the novel approach to an approximate scale symmetry breaking up to the phase transition is suggested. 

We proposed the combined method to match the effective model of the dilaton in terms of the glueball to the one of free gluons at the PT.
%The possible determination of the phase boundary between the confinement - deconfinement border and the high $T$ plasma phase can be seen inside the conformal window.
We find the PT is achieved at high chemical potential $\mu$ (the case B) with smaller particle momentum (and, hence, the energy). 
%The fluctuation of the particle density increases indefinitely with high temperatureIn the vicinity of the PT one has the scalar condensate with 
There is the sharp increasing of the fluctuation of the particle density at high ratio $T/\mu\sim \ln^{-1} (1/\bar\mu_{0})$ (\ref{e27}). 
%The fluctuation of the particle density disappears at the critical value of the temperature (\ref{e28}).
%At the critical temperature the fluctuation of the particle density disappears ({\refe28}).
%The boson stars are unstable and the showers of the photons can be registered because of unstable dilatons (glueballs) decaying to primary photons. These decays could contribute to new sources of cosmic rays.  
%The importance of the BS is actual and evident, because the boson stars hide a scalar part of DM from direct observation. After the formation and becoming large in size, the BS may explode into leptons via decays of the dark %photons [36] or may emit  the primary photons which  may explain the conformal anomaly in EW and EM hidden sectors. These could contribute to new sources of cosmic rays.
In the laboratory experiments,  the PT can be found as those followed by the IRFP where the primary photons are detected. The origin of these photons is the CA via the decays of the dilatons. When the incident energy scans from high to lower values, one can find the non-monotonous behaviour in fluctuations of  primary photons: 
%(primary photons production compared to that from  $\pi^{0}\rightarrow\gamma\gamma $ decay). 
%The latter  will indicate the approaching of CP. 
the fluctuations rate $r_{\gamma\gamma}$ grows up in the IR to become very large at the PT (\ref{e209}). 
%At the critical temperature the fluctuation of the particle density disappears (\ref{e28}).
Both the PT and the CP have the very clear signature: the shower increasing of the photons flow in the detector compared to that produced by light hadrons. 
The information about the event with the PT for the given experimental conditions can be obtained by measuring the ratios of $\gamma$-quanta yields and compared (fitting) to known model with $T$ and $\mu$.

%The boson stars are unstable and the showers of the photons can be registrated because of unstable dilatons (glueballs) decaying to primary photons. These decays could contribute to new sources of cosmic rays.  

\end{document}